\documentclass{JHEP3} 

\usepackage{epsfig,multicol,bbm}
\usepackage{cite}

\newcommand\fverb{\setbox\fverbbox=\hbox\bgroup\verb}
\newcommand\fverbdo{\egroup\medskip\noindent%
            \fbox{\unhbox\fverbbox}\ }
\newcommand\fverbit{\egroup\item[\fbox{\unhbox\fverbbox}]}
\newbox\fverbbox

\title{Seesaw Type I and III at the LHeC}

\author{Liang Han$^a$, Xiao-Gang He$^{b,c}$, Ma Wen-Gan$^a$, Wang Shao-Ming$^a$, and Zhang Ren-You$^a$\\
$^a$Department of Modern Physics, University of Science and Technology of China, Hefei, Anhui 230026 \\
$^b$INPAC and Department of Physics, Shanghai Jiao Tong University, Shanghai, 200240\\
$^c$Department of Physics and Center for Theoretical Sciences, National Taiwan University, Taipei\\
E-mail: \email{hanl@ustc.edu.cn}, \email{hexg@phys.ntu.edu.tw},
\email{mawg@ustc.edu.cn}, \email{smwang@mail.ustc.edu.cn},
\email{zhangry@ustc.edu.cn}}

\abstract{We study the potential of testing the seesaw type-I and
III models at the Large Hadron-electron Collider (LHeC), an e-p
collision mode at the CERN collider. The e-p collision mode provides
an excellent place to study lepton number violating processes $e^-
p\to N_{i} j +X \to e^+ W^-j+ X$, $e^- p\to N_{i} j +X \to
\tau^{\pm}W^{\mp}j+ X$ and $e^- p\to E_{i} j +X \to \tau^-Zj+ X$
with $W$ and $Z$ into hadron jets. Here $N_{1,2,3}$ and $E_{1,2,3}$
are heavy Majorana neutrinos and heavy charged leptons, and $j$ is a
hard hadron jet. Although the process $e^- p\to N_{i} j +X \to e^+
W^-j+ X$ is stringently constrained from neutrinoless double-beta
decay, there are solutions where this constraint can be satisfied
with sizeable production cross section. With the electron energy
$E_e = 140$ GeV and proton energy $E_p = 7$ TeV, we find that the
cross section for the heavy charged lepton $E$ production can reach
a few $fb$ when the heavy charged lepton mass $m_E < 600$ GeV. For
the heavy neutrino $N$ production, the cross section can be as large
as a few $fb$ for the mass scale as high as 1 TeV, higher than what
can be achieved by the p-p collision mode of LHC with the same
related heavy neutrino couplings.  }


\begin{document}

\section{Introduction}

Various experiments have now established that neutrinos have masses
and mixes with each other~\cite{pdg}. The masslessness of the
neutrinos in the minimal standard model (SM) implies that one has to
go beyond it to account for this observation. Among a number of
possibilities that have been proposed, the most popular ones are the
seesaw scenarios~ \cite{Konetschny:1977no,Schechter:1980nm,
Cheng:1980nm,Zee:1980at,Mohapatra:1981nm,Lazarides:1981pl,Bahu:1988mo,Ma:1998ptn,
Pilaftsis:1992ri,Minkowski:1977sc,Yanagida:1979po,GellMann:1979su,
Glashow:1980po,Mohapatra:1979ia,Keung:1983uu,seesawIII} in which new
particles are introduced with masses sufficiently large to make the
neutrino masses small.

In the so-called seesaw type-I (ST-I) and type-III (ST-III)
models~\cite{Minkowski:1977sc,Yanagida:1979po,GellMann:1979su,
Glashow:1980po,Mohapatra:1979ia,Keung:1983uu,seesawIII}, the heavy
particles responsible for giving masses to the light neutrinos are
neutral fermions -- the heavy neutrinos. The best way to test seesaw
models is to produce the heavy neutrino $N$, as well as their
charged partners in the case of the ST-III model.
Several studies of seesaw models at the LHC in p-p collision mode
have been carried out~\cite{Han:2006sf,Atre:2009ts,delAguila:2006np,
delAguila:2007emh,Kersten:2007vk,BarShalom:2008gt,Perez:2009to,Chun:2003th,
Han:2005nm,Chen:2007tr,Hector:2007tn,Han:2007pr,Garayoa:2008nm,Kadastik:2008ddo,
Akeroyd:2008pm,Chao:2008ac,Chao:2008mq,Perez:2008ta,Perez:2008nm,Perez:2009la,
delAguila:2008cj,sst3,Bajc:2007zf,Perez:2009ts,Li:2009nm,Huitu:2008sf,Arhrib:2009mz}.
It has been shown that at the LHC single production of heavy seesaw
particle with mass of order a few hundred GeV can be
probed~\cite{delAguila:2008cj}. For the ST-III model, pair
production of heavy seesaw particle up to 1 TeV can be
achieved~\cite{delAguila:2008cj,sst3,Bajc:2007zf,Perez:2009ts,Li:2009nm}.

In this work we study the possibility of testing the ST-I and ST-III
models at the Large Hadron-electron Collider (LHeC), an possible
operation mode with e-p collision for the LHC. The e-p collision
mode provides an excellent place to study lepton number violating
processes $e^- p\to N_{i} j +X \to e^+ W^-j+ X$, $e^- p\to N_{i} j
+X \to \tau^{\pm}W^{\mp}j+ X$ and $e^- p\to E_{i} j +X \to \tau^-Zj+
X$ with $W$ and $Z$ into hadron jets. Although there is stringent
constraint from neutrinoless double-beta decay, it only constrains
the $e^- p\to N_{i} j +X \to e^+ W^-j+ X$ process. Even for the $e^-
p\to N_{i} j +X \to e^+ W^-j+ X$ process, there are solutions with
non-degenerate heavy neutrinos which can satisfy the constraint from
neutrinoless double-beta decay. Then the cross sections for the
processes $e^-p \to N_ij+X \to e^+W^+j+ X$ ($i =1,\;2,\;3$) can be
sizeable. With the electron energy $E_e = 140$ GeV and proton energy
$E_p = 7$ TeV, we find that the cross section for the heavy charged
lepton $E$ production can reach a few $fb$ when the heavy charged
lepton mass $m_E < 600$ GeV. For the heavy neutrino $N$ production,
the cross section can be as large as a few $fb$ for the mass scale
as high as 1 TeV, higher than what can be achieved by the p-p
collision mode of LHC with the same related heavy neutrino
couplings.

\section{Theoretical Models and Calculations}

\subsection{The Seesaw Type-I Model}

In the ST-I model, the seesaw mechanism is realized by introducing
right-handed neutrinos that are singlets under the SM gauge group
$SU(3)_C\times SU(2)_L\times U(1)_Y$ and can therefore have Majorana
masses~\cite{Minkowski:1977sc,Yanagida:1979po,GellMann:1979su,
Glashow:1980po,Mohapatra:1979ia,Keung:1983uu}. We assume for
definiteness that there are three of these heavy neutrinos,
$N_{iR}$, responsible for giving masses to the light
neutrinos,~$\nu_{iL}^{}$. The Lagrangian describing the masses of
the neutrinos can be written as
\begin{eqnarray}
{\cal L} \,\,=\,\,
- \bar N_{iR}^{} (Y_D^{})_{ij}^{} \tilde H^\dagger L_{jL}^{}
- \mbox{$\frac{1}{2}$} \bar N_{iR}^{} (M_N^{})_{ij}^{} N_{jR}^c
\,\,+\,\, {\rm H.c.} ~,
\end{eqnarray}
where summation over \,$i,j=1,2,3$\, is implied.  $Y_D^{}$ is the
3$\times$3 Yukawa coupling matrix. $\tilde H=i\tau_2^{}H^*$ with
$\tau_2^{}$ being the usual Pauli matrix.
$H=\bigl(\phi^+,~~(v+h+i\eta)/\sqrt2\bigr){}^T$ is the Higgs doublet
and $v$~its vacuum expectation value. $L_{iL}^{}
=\bigl(\nu_{iL}^{}~~l^-_{iL}\bigr){}^T$ is the left-handed lepton
doublet. $M_N$~is the Majorana mass matrix, and $N_{iR}^c$\, is the
charge conjugation of $N_{iR}$.

In the ($\nu_L^{}, N_R^{c})^T$ basis the mass matrix is given by
\begin{eqnarray}
M_{\rm seesaw}^{} \,\,=\,\, \left( \begin{array}{cc} 0 & m_D^T
\vspace{0.5ex} \\ m_D^{} & M_N^{}
\end{array} \right) ,
\end{eqnarray}
with the Dirac mass matrix  \,$m_D^{}=v Y_D^{}/\sqrt2$.\,

One can relate the weak eigenstates $\nu_{iL}^{}$ and $N_{iR}^c$ to the corresponding
mass eigenstates according to
\begin{eqnarray} \label{basis}
\left( \begin{array}{c} \nu_L^{} \vspace{1ex} \\ N_R^c \end{array} \right)
=\,\, U \left( \begin{array}{c} \nu_{mL}^{} \vspace{1ex} \\ N_{mL}^{} \end{array} \right) ,
\hspace{5ex}
U \,\,\equiv\,\, \left(\begin{array}{cc} V_{\nu\nu}^{} & V_{\nu N}^{} \vspace{1ex} \\
V_{N \nu}^{} & V_{NN}^{} \end{array}\right) .
\end{eqnarray}
where  $\nu_{mL}^{}$ and $N_{mL}^{}$ are column matrices containing the mass eigenstates.
Thus $U$ is unitary and diagonalizes $M_{\rm seesaw}$,
\begin{eqnarray} \label{mdiag}
\left(\begin{array}{cc} \hat m_\nu^{} & 0 \vspace{0.5ex} \\ 0 & \hat
M_N^{} \end{array}\right) =\,\, U^{T}  M_{\rm seesaw}^{} U ~,
\end{eqnarray}
where  \,$\hat m_\nu^{}={\rm
diag}\bigl(m_{\nu_1},m_{\nu_2},m_{\nu_3}\bigr)$\,  and \,$\hat
M_N^{} = {\rm diag}\bigl(m_{N_1}, m_{N_2}, m_{N_3}\bigr)$.\, Note that the submatrices  $V_{\nu\nu}$, $V_{\nu N}$, $V_{N \nu}$,
and $V_{NN}$ are not unitary.

In the mass eigenstate basis, dropping the subscript $m$ for mass eigenstate fields,
the gauge and Higgs interactions with $N$ are given by
\begin{eqnarray}
&&\mathcal{L}_{CC}=
- \frac{g}{\sqrt2} \Bigl(
\bar l_L^{}\, \gamma^\mu V_{PMNS}^{} \nu_{L}^{} W^-_\mu +
\bar l_L^{}\, \gamma^\mu V_{l N}^{} N_{L} ^{} W^-_\mu \,+\, {\rm H.c.} \Bigr)
\nonumber \\
&&\mathcal{L}_{NC}=
\frac{g}{2 c_{\rm w}^{}} \Bigl(
\bar\nu_{L}^{}\, \gamma^\mu V^\dagger_{PMNS} V_{PMNS}^{} \nu_{L}^{}
+ \bar N_{L}^{}\, \gamma^\mu V_{l N}^\dagger V_{PMNS}^{} \nu_{L}^{}
\nonumber \\ && \hspace*{8ex} +\,\,
\bar\nu_{L}^{}\, \gamma^\mu V^\dagger_{PMNS} V_{l N}^{} N_{L}^{}
+ \bar N_{L}^{}\, \gamma^\mu V_{l N}^\dagger V_{l N}^{} N_{L}^{} \Bigr) Z_\mu^{}
\nonumber\\
&&\mathcal{L}_{S}=\,\,\frac{g}{2M_W} \Bigl[
\overline{\nu_{L}^{\!c}}\,\hat m_\nu^{}V_{PMNS}^\dagger
V_{PMNS}^{}\nu_{L}^{} + \overline{N_{L}^{\!c}}\,\hat M_N^{} V_{l
N}^\dagger V_{PMNS}^{} \nu_{L}^{} \nonumber \\ && \hspace*{4ex}
+\,\, \overline{\nu_{L}^{\!c}}\,\hat m_\nu^{} V_{PMNS}^\dagger V_{l
N}^{} N_{L}^{} + \overline{N_{L}^{\!c}}\,\hat M_N^{} V_{l N}^\dagger
V_{l N}^{} N_{L}^{} \,+\, {\rm H.c.} \Bigr] {h} \label{st-i-int}
\end{eqnarray}
where $g$ is the $SU(2)_L$ coupling and $c_W = \cos\theta_W$ with
$\theta_W$ being the Weinberg angle. $V_{lN} = V_{\nu N}$. The
Pontecorvo-Maki-Nakagawa-Sakata (PMNS) matrix $V_{PMNS} =
V_{\nu\nu}$ is the usual light neutrino mixing matrix~\cite{Maki:1962mu,Pontecorvo:1967fh}.
Strictly speaking it is not unitary in seesaw models. But to the
leading order the non-unitary effects can be neglected.

It is clear from the above expressions that the heavy
neutrinos $N_R$ can interact with the SM gauge bosons via mixing at
the tree level. The leading parton level single heavy seesaw particle production at e-p collision is through
\begin{eqnarray}
e + q \to N + q'~~(\mbox{through t-channel $W$-exchange.})\;.
\end{eqnarray}

Combining  the interaction Lagrangian in Eq.(\ref{st-i-int}) and the
charged quark current interaction,
\begin{eqnarray}
{\cal L}' &\!\!=&\!\!
-\frac{g}{\sqrt2} \Bigl(
\bar u\, \gamma^\mu V_{CKM} d W^+_\mu \, +  {\rm H.c.} \Bigr)\;,
\end{eqnarray}
where $V_{CKM}$ is the Cabbibo-Kobayashi-Maskawa (CKM) mixing
matrix~\cite{Cabbibo:1963us,Kobayashi:1973cp} for quarks, we obtain the corresponding
amplitudes, for example, ${\cal M}$ for $e^-(p_1) + u(p_2) \to
N_i(p_3) +d(p_4)$,
\begin{eqnarray}
{\cal M} &=& i {g^2\over 2}(V_{lN})_{1i}^* V^*_{ud} {g_{\mu\nu}\over \hat{t} - m^2_W } \nonumber \\
& \times & \left[ \bar N_i \gamma^\mu P_L e \bar d \gamma^\nu P_L u
+\frac{1}{m_W} \bar N_i (m_{N_i} P_L - m_e P_R) e \bar d (m_d P_L -
m_u P_R) u \right ]\;,
\end{eqnarray}
where $\hat{t} = (p_1 - p_3)^2$. We obtain the t-channel
$W$-exchange cross section for the partonic process $e^-(p_1) +
q(p_2) \to N_i(p_3) +q'(p_4)$ expressed as
\begin{eqnarray}
\sigma_W(e^-q \to N_i q') &=& \int_{\hat{t}_0}^{\hat{t}_1}
\mathrm{d} \hat{t} \; \frac{g^4|(V_{lN})_{1i}|^2}
{64\pi\hat s p_{in}^2}\; \frac{1}{(\hat{t}-m_W^2)^2} \nonumber \\
&\times& \left\{ \begin{array}{ll}
|V_{qq'}|^2(p_1 \cdot p_2)(p_3 \cdot p_4), & \textrm{if $q=u,c,q'=d,s,b$.} \\
|V_{\bar{q}^{'}\bar q}|^2(p_1 \cdot p_4)(p_2 \cdot p_3), & \textrm{if $q=\bar d,\bar s, q'=\bar u,\bar c$.}
\end{array} \right.
\label{sigma-w}
\end{eqnarray}
where
\begin{eqnarray}
\hat s &=& (p_1+p_2)^2, \nonumber \\
p_{in}^2 &=& \frac{(\hat{s}+m_2^2-m_1^2)^2}{4\hat s}-m_2^2, \nonumber \\
p_{out}^2 &=& \frac{(\hat{s}+m_4^2-m_3^2)^2}{4\hat s}-m_4^2,  \nonumber \\
\hat{t}_0 &=& \frac{(m_1^2-m_2^2-m_3^2+m_4^2)^2}{4\hat s}-(p_{in}+p_{out})^2, \nonumber \\
\hat{t}_1 &=& \frac{(m_1^2-m_2^2-m_3^2+m_4^2)^2}{4\hat
s}-(p_{in}-p_{out})^2. \label{sigma-para-1}
\end{eqnarray}

\subsection{The Seesaw Type-III Model}
The ST-III model consists of, in addition to the SM particles,
left-handed triplet leptons with zero hypercharge, $\Sigma_L\sim
(1,3,0)$ under $SU(3)_C\times SU(2)_L\times
U(1)_Y$~\cite{seesawIII}. We write the component fields as
\begin{eqnarray}
&&\Sigma_L=
\left(
  \begin{array}{cc}
    \Sigma_L^0/\sqrt{2} & \Sigma^+_L \\
    \Sigma_L^- & -\Sigma_L^0/\sqrt{2} \\
  \end{array}
\right).
\end{eqnarray}
The charge conjugated form is
\begin{eqnarray}
\ \ \Sigma^c_L= \left(
  \begin{array}{cc}
    \Sigma_L^{0c}/\sqrt{2} & \Sigma^{-c}_L \\
    \Sigma_L^{+c} & -\Sigma_L^{0c}/\sqrt{2} \\
  \end{array}
\right).
\end{eqnarray}
Note that $\Sigma^c_L $ is right-handed field.

The renormalizable Lagrangian involving $\Sigma_L~(\Sigma_L^c)$ is
given by
\begin{eqnarray}
\mathcal{L}&=&{\rm Tr}[\overline{\Sigma_L}i\gamma_\mu
{D^\mu}\Sigma_L]-{1\over 2}{\rm Tr}[\overline{\Sigma_L^c}M_\Sigma
\Sigma_L+\overline{\Sigma_L}M_\Sigma^\ast
\Sigma_L^c]-\overline{L_L}\sqrt{2}Y_\Sigma^\dagger \Sigma_L^c
\tilde{H}-\tilde{H}^\dagger \overline{\Sigma_L^c}\sqrt{2}Y_\Sigma
L_L\;. \nonumber \\
\end{eqnarray}

One can define $\Psi\equiv \Sigma_L^-+\Sigma_L^{+c}$ with
$\Psi_L=\Sigma_L^-$, $\Psi_R=\Sigma_L^{+c}$ to obtain the charged
and neutral  lepton mass matrices in the basis, $(l_{L},
\Psi_L)^T$ and $(\nu_L, \Sigma^0_L)^T$, respectively,  as
\begin{eqnarray}
\left(
  \begin{array}{cc}
    m_l & 0 \\
    v Y_\Sigma & M_\Sigma \\
  \end{array}
\right)\;,\;\;
 \left(
  \begin{array}{cc}
    0 & vY_\Sigma^T/2\sqrt{2} \\
    vY_\Sigma /2\sqrt{2} & M_\Sigma/2 \\
  \end{array}
\right)\;.
\end{eqnarray}
The diagonalization of the mass matrices can be achieved by making
unitary transformations on the triplet, the charged and neutral
leptons defined in the following
\begin{eqnarray}
&&\left(
    \begin{array}{c}
      l_{L,R} \\
      \Psi_{L,R} \\
    \end{array}
  \right)
  =U_{L,R}
  \left(
    \begin{array}{c}
      l_{mL,R} \\
      E_{mL,R} \\
    \end{array}
  \right), \ \ \
\left(
    \begin{array}{c}
      \nu_{L} \\
      \Sigma^0_{L} \\
    \end{array}
  \right)
  =U_0
  \left(
    \begin{array}{c}
      \nu_{mL} \\
      N_{mL} \\
    \end{array}
  \right),
\end{eqnarray}
where $U_{L,R}$ and $U_0$ are $6\times6$ unitary matrices which we decompose
them into $3\times 3$ block matrices as
\begin{eqnarray}
&&U_L\equiv
\left(
  \begin{array}{cc}
    U_{Lll} & U_{Ll\Psi} \\
    U_{L\Psi l} & U_{L\Psi\Psi} \\
  \end{array}
\right), \ \ U_R\equiv \left(
  \begin{array}{cc}
    U_{Rll} & U_{Rl\Psi} \\
    U_{R\Psi l} & U_{R\Psi\Psi} \\
  \end{array}
\right), \ \ U_0\equiv \left(
  \begin{array}{cc}
    U_{0\nu\nu} & U_{0\nu\Sigma} \\
    U_{0\Sigma \nu} & U_{0\Sigma\Sigma} \\
  \end{array}
\right).
\end{eqnarray}

To the leading order, we have gauge and Higgs interaction terms
involving heavy triplet leptons as~\cite{sst3,Bajc:2007zf,Perez:2009ts,Li:2009nm}
\begin{eqnarray}
&&\mathcal{L}_{NC(A+Z)}=e\overline{E}\gamma^\mu E A_\mu + gc_W\overline{E}
\gamma^\mu E Z_\mu \;,\nonumber\\
&&\mathcal{L}_{NCZ}={g\over
2c_W}[\overline{\nu}(V_{PMNS}^\dagger V_{lN}\gamma^\mu
P_L-V_{PMNS}^TV^{\ast}_{lN}\gamma^\mu P_R)N_{}
+\sqrt{2}\overline{l}V_{lN}\gamma^\mu
P_LE_{}+ {\rm H.c.}]Z_\mu\;,\nonumber\\
&&\mathcal{L}_{CC}=-g [ \overline{E}\gamma^\mu N +{1\over
\sqrt{2}}\overline{l}V_{lN}\gamma^\mu P_LN_{} +
\overline{E}V_{lN}^TV_{PMNS}^\ast\gamma^\mu
P_R\nu_{}]W^-_\mu+{\rm H.c.}\;,\label{st-iii-int}\\
&&\mathcal{L}_{S}=-{g\over 2M_W}[\overline{\nu}(V_{PMNS}^\dagger
V_{lN}\hat M_{N}P_R+V_{PMNS}^TV^{\ast}_{lN}\hat M_{N}P_L)N_{}+
\sqrt{2}\overline{l}V_{lN}\hat M_{E}P_RE_{}]h+{\rm H.c.}\;,\nonumber
\end{eqnarray}
with $V_{lN}\equiv V_{l\Sigma}^L = -Y^\dagger_\Sigma v
M^{-1}_\Sigma/\sqrt{2}$. In the above, all fields are in mass
eigenstates. $\hat M_{E},\hat M_{N}$ are eigen-mass matrices of
$E$, $N$.

In this model, there is also the possibility of producing the
heavy charged leptons, $E$. The heavy seesaw particle production
at the e-p collider involves the following partonic processes,
\begin{eqnarray}
&&e^-q \to N q'~~(\mbox{through t-channel $W$-exchange.})\;, \nonumber\\
&&e^- q \to E^- q~~(\mbox{through t-channel $Z$- and
$h$-exchange.})\label{ggg}
\end{eqnarray}
The cross section, $\sigma_W( e^- q \to N q')$, has the same
expression as that given in Eq.(\ref{sigma-w}). The t-channel $Z$-
and $h$-exchange for $E$ production cross section, $\sigma_{Zh}(e^-
q \to E^- q)$, can be obtained using the interactions in
Eq.(\ref{st-iii-int}). The contribution from $h$-exchange involves a
small light quark-Higgs Yukawa coupling and therefore is small. For
the $Z$-exchange contribution, we denote the processes as $e^-(p_1)
+ q(p_2) \to E_i^-(p_3) + q(p_4)$ and have
\begin{eqnarray}
\sigma_{Zh}(e^- q \to E_i^- q) = \int_{\hat{t}_0}^{\hat{t}_1}
\mathrm{d} \hat{t} \; \frac{g^4|(V_{lN})_{1i}|^2}{32\pi c_W^4\hat s
p_{in}^2}\; \frac{1}{(\hat{t}-m_Z^2)^2}
~~~~~~~~~~~~~~~~~~~~~~~~ \nonumber \\
\times \left\{ \begin{array}{ll}
\left [ g_L^2(q)(p_1 \cdot p_2)(p_3 \cdot p_4)+g_R^2(q)(p_1 \cdot p_4)(p_2 \cdot p_3)\right ],
& \textrm{if $q=u,c,d,s$.} \\
\left [ g_R^2(q)(p_1 \cdot p_2)(p_3 \cdot p_4)+g_L^2(q)(p_1 \cdot
p_4)(p_2 \cdot p_3)\right ], & \textrm{if $q=\bar u,\bar c, \bar d,
\bar s$.}
\end{array} \right.
\label{sigma-zh}
\end{eqnarray}
where $\hat s$, $\hat{t}$, $p_{in}$, $\hat{t}_0$ and $\hat{t}_1$
have the same definitions as in
Eqs.(\ref{sigma-w})-(\ref{sigma-para-1}), and
\begin{eqnarray}
g_L(u)&=&g_L(c)=g_L(\bar u)=g_L(\bar c)=\frac{1}{2}-\frac{2}{3}s_W^2,\nonumber \\
g_R(u)&=&g_R(c)=g_R(\bar u)=g_R(\bar c)=-\frac{2}{3}s_W^2, \nonumber \\
g_L(d)&=&g_L(s)=g_L(\bar d)=g_L(\bar s)=-\frac{1}{2}+\frac{1}{3}s_W^2, \nonumber \\
g_R(d)&=&g_R(s)=g_R(\bar d)=g_R(\bar s)=\frac{1}{3}s_W^2.
\end{eqnarray}

\section{Numerical analysis}

\subsection{Production of $N$ and $E$ at the LHeC}

To obtain the cross sections for $e^-p \to N_{1} j+ X$ process ($j$
indicates a hard quark jet) in the ST-I and ST-III models,  and for
the $e^-p \to E^-_1 j+X$ process in the ST-III model, one needs to
fold in the parton distributions (PDFs) of the proton.  We use
CTEQ6L1 PDFs for the LO calculations.

The calculations for the processes $e^- p \to N_{1} j+ X$ and $e^-p
\to E^-_{1} j + X$ involve the contributions of the partonic
processes $e^{-}q \to N_1 q'$ (where $q=u,d,c,s$ and
$q'=u,d,s,c,b$), and $e^{-}q \to E^-_1 q$ (where $q=u,d,c,s$),
respectively. We neglect the masses of electron and light quarks
$u,d,s$, and assume $m_h=115$ GeV. For the CKM quark mixing
parameters, we will use central values $\lambda = 0.2257$, $A =
0.814$, $\bar \rho = 0.135$ and $\bar \eta = 0.349$ in
Ref.\cite{pdg} to determine the angles and phase in the Particle
Data Group (PDG) parametrization of CKM matrix and use them in our
numerical calculations. We have, $s_{12} = 0.2257$, $s_{23} =
0.0415$, $s_{13} = 0.00359$ and $\delta = 68.88^\circ$. The other
input parameters involved are taken as~\cite{pdg}:
\begin{eqnarray}
\alpha^{-1}(m_Z^2)&=&127.918,~~~m_Z=91.1876~{\rm GeV},~~~m_W=80.398~{\rm GeV}, \nonumber\\
m_c&=&1.3~{\rm GeV},~~~m_b=4.2~{\rm GeV}.
\end{eqnarray}

\par
As a demonstration of numerical results we take $E_p=7~{\rm TeV}$,
$E_{e}=70~{\rm GeV}$ or $140~{\rm GeV}$, and set the factorization
scales $\mu_f=\frac{1}{2}m_{N_1}$ for process $e^- p \to N_1 j+X$,
and $\mu_f=\frac{1}{2}m_{E_1}$ for process $e^- p \to E_1^- j+X$.

\begin{table}[tp]
\begin{center}
\begin{tabular}{|c|c|c|c|c|} \hline
 $m_{N_1}(m_{E_1})$& \multicolumn{2}{c|}{$\sigma(e^{-} p \to N_1 j+X)/|(V_{lN})_{11}|^2 ~(pb)$ } &
     \multicolumn{2}{c|}{$\sigma(e^{-}p \to E^-_1 j+X)/|(V_{lN})_{11}|^2 ~(pb)$ } \\ \cline{2-5}
  $({\rm GeV})$& $E_e=70{\rm GeV}$ & $E_e=140{\rm GeV}$ & $E_e=70{\rm GeV}$ & $E_e=140{\rm GeV}$   \\ \hline
 100 & 147.8 & 211.1 & 103.4  &153.7  \\
 200 & 75.15  & 117.2 & 49.71 &81.89  \\
 400 & 22.67  & 46.29 & 13.46 &29.63 \\
 600 & 6.049  & 19.46 & 3.27 &11.55 \\
 800 & 1.125  & 7.614 & 0.552 &4.22 \\
 1000& 0.1028 & 2.549 & 0.0448 &1.32 \\
 \hline
\end{tabular}
\end{center}
\begin{center}
\begin{minipage}{14cm}
\caption{\label{tab1} The numerical results for the cross sections
for $e^- p \to N_{1}j+X$ in the ST-I and ST-III models, and $e^- p
\to E^-_{1} j+X$ in the ST-III model at the LHeC as the functions of
$m_{N_1}$ and $m_{E_1}$, respectively. There we take $E_p=7~{\rm
TeV}$. }
\end{minipage}
\end{center}
\end{table}

\par
\begin{figure}[htbp]
\vspace{5mm}
 \centering
 \includegraphics[width=10cm]{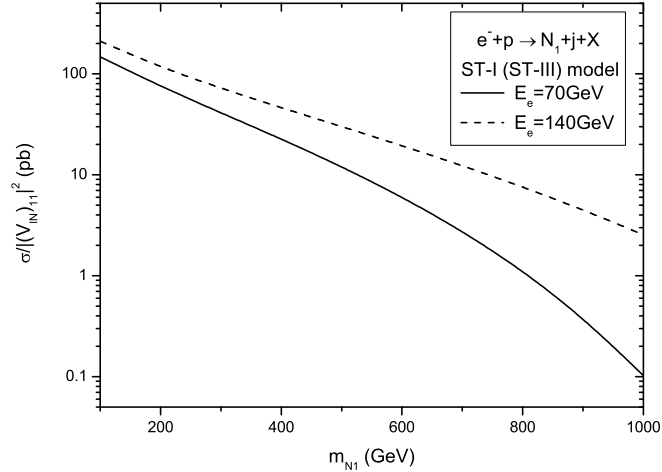}
\caption{\label{fig1} The cross sections for $N_1j$ production at
the LHeC as the functions of $m_{N_1}$ with $E_p=7~{\rm TeV}$, where
the full-line is for $E_e=70~{\rm GeV}$ and the dashed-line for
$E_e=140~{\rm GeV}$. }
\end{figure}
\begin{figure}[htbp]
\vspace{5mm}
 \centering
 \includegraphics[width=10cm]{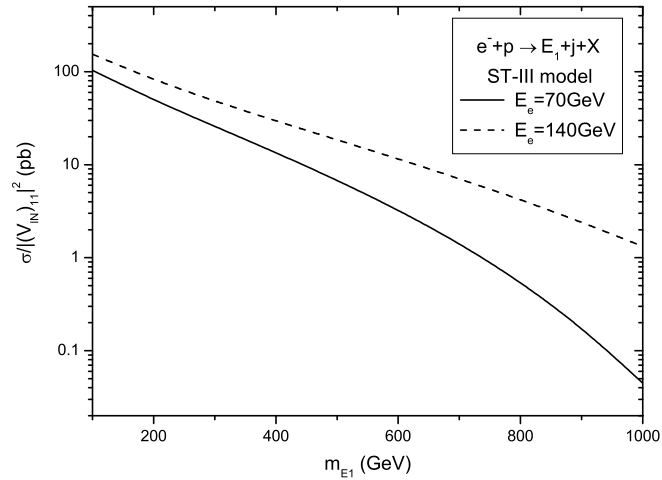}
\caption{\label{fig2} The cross sections for $E^-_1j$ production at
the LHeC as the functions of $m_{E_1}$ with $E_p=7~{\rm TeV}$, where
the full-line is for $E_e=70~{\rm GeV}$ and the dashed-line for
$E_e=140~{\rm GeV}$.  }
\end{figure}

In Fig.\ref{fig1} and Fig.\ref{fig2}, we depict the production rates
of the processes $e^- p \to N_1j + X$ in the ST-I and ST-III models
and $e^- p \to E^-_1 j + X$ in the ST-III model as functions of
the masses of $N_1$ and $E^-_1$, respectively. For the $N_1j$ production
process the $\sigma(e^{-} p \to N_1 j+X)/|(V_{lN})_{11}|^2$ results
in both the ST-I and ST-III models are the same. In Table \ref{tab1}
we also list some of the representative numerical results which are
read off from Fig.\ref{fig1} and Fig.\ref{fig2}.


Whether it is possible at the LHeC to test the ST-I and ST-III
models not only depends on the masses of the heavy states, but also
crucially depends on how large $V_{lN}$ can be in order to have
sufficient number of $N_i$ and $E^-_i$ to be produced. If there is
only one generation, one would obtain $V_{l N}$ of order $m_D/m_N$
whose magnitude is $\sqrt{|m_\nu/m_N|}$. With light neutrino mass
constrained to be less than of order $1~{\rm eV}$, the magnitude of
mixing $V_{l N}$ is bounded by $10^{-6}\sqrt{100\mbox{GeV}/m_N}$.
Even with $m_N$ of order $100~{\rm GeV}$, the mixing is extremely
small. With such a small mixing it is not possible to produce enough
number of heavy neutrinos to study its properties at the LHeC.
However, this conclusion is true only for one generation of
neutrino. With more than one generation, one can evade the
constraint $|V_{l N}| = \sqrt{|m_\nu/m_N|}$. There are solutions of
the form $V_{l N} = V_0 + V_\delta$ satisfying constraints on light
neutrino masses and mixing with some of the elements in $V_0$ to be
left unconstrained from neutrino
masses~\cite{Kersten:2007vk,Buchmuller:1990du,Buchmuller:1991tu,Gluza:2002vs,Ingelman:1993ve,
Pilaftsis:2004xx,Pilaftsis:2005rv,Ma:2009du,Xing:2010na,He:2009xd,He:2009lm},
as long as $V_0 \hat{M}_N V_0^T = 0$ which requires that $V_0$ being
rank one~\cite{He:2009lm}. Such solutions have interesting
implications~\cite{large1,Park:2010nm,Chen:2010eo}. Constraints on
the sizes of elements in $\epsilon = V_{lN}V^\dagger_{lN}$ then come
from flavor changing neutral current processes. For the ST-I model,
we have the constraints
as~\cite{delAguila:2008cj,ewpd,delAguila:2008ks}
\begin{eqnarray} \label{eii}
\epsilon_{11}^{} \,\,\le\,\, 3.0\times10^{-3} ~, \hspace{5ex}
\epsilon_{22}^{} \,\,\le\,\, 3.2\times10^{-3} ~, \hspace{5ex}
\epsilon_{33}^{} \,\,\le\,\, 6.2\times10^{-3} ~,
\end{eqnarray}
and Refs.\cite{ewpd,delAguila:2008ks,typeI_fcnc} give the limitations as
\begin{eqnarray} \label{eij}
|\epsilon_{12}^{}| \,\,\le\,\, 1\times10^{-4} ~, \hspace{5ex}
|\epsilon_{13}^{}| \,\,\le\,\, 0.01 ~, \hspace{5ex} |\epsilon_{23}^{}| \,\,\le\,\, 0.01 ~.
\end{eqnarray}
For heavy neutrinos coupling to the electron, neutrinoless double-beta decay
imposes~\cite{Belanger:1995nh}
\begin{eqnarray} \label{ee0nu}
\biggl|\sum_{i=1}^3(V_{l N})_{1i}^2/m_{N_i}^{}\biggr| \,\,\le\,\,
5\times10^{-8}{\rm\,GeV}^{-1} ~.
\end{eqnarray}
Finally, there are also constraints from searches for SM-singlet
neutrinos by the L3 and DELPHI experiments at LEP~\cite{lep,Abreu:1996pa} on the
individual elements $(V_{l N})_{2i}$ and $(V_{lN})_{3i}$ which may
be stronger than those inferred from Eqs.~(\ref{eii})
and~(\ref{eij}), depending on $m_{N_i}$.

For the ST-III model, because additional $E^\pm$ exist in the model
the constraints are slightly stronger. They
are~\cite{delAguila:2008cj,ewpd,delAguila:2008ks}
\begin{eqnarray} \label{eii'}
\epsilon_{11}^{} \,\,\le\,\, 3.6\times10^{-4} ~, \hspace{5ex}
\epsilon_{22}^{} \,\,\le\,\, 2.9\times10^{-4} ~, \hspace{5ex}
\epsilon_{33}^{} \,\,\le\,\, 7.3\times10^{-4} ~,
\end{eqnarray}
whereas from the measurements of lepton-flavor violating transitions~\cite{Abada:2008ea}
\begin{eqnarray} \label{fcnc'}
|\epsilon_{12}^{}| \,\,\le\,\, 1.7\times10^{-7} ~, \hspace{5ex}
|\epsilon_{13}^{}| \,\,\le\,\, 4.2\times 10^{-4} ~, \hspace{5ex}
|\epsilon_{23}^{}| \,\,\le\,\, 4.9\times 10^{-4} ~.
\end{eqnarray}
In addition, direct searches for heavy charged leptons at colliders
impose constraints on the mass of $E$, and hence the mass of $N$ as
well, namely
\,$m_{N_i},m_{E_i}\,\raisebox{-0.4ex}{\footnotesize$\stackrel{\textstyle>}{\sim}$}\,100$\,GeV.\,

If the heavy neutrinos are nearly degenerate, we can get the
relation for the cross sections from Eq.(\ref{sigma-w}) expressed as
\begin{eqnarray} \label{sigma-ratio}
&&\sigma(e^- p \to N_1 j+X):\sigma(e^- p \to N_2 j+X):\sigma(e^- p
\to N_3 j+X) \nonumber \\
&&=|(V_{lN})_{11}|^2:|(V_{lN})_{12}|^2:|(V_{lN})_{13}|^2,
\end{eqnarray}
From the definition of $\epsilon$ matrix we have $\epsilon_{11} =
\sum_{i=1}^3 |(V_{lN})_{1i}|^2$. Then we obtain
\begin{eqnarray} \label{sum-sigma}
\sum_{i=1}^3 \sigma(e^- p \to N_i j+X) &=& \sum_{i=1}^3
|(V_{lN})_{1i}|^2 \sigma(e^- p \to N_1 j+X)/|(V_{lN})_{11}|^2 \nonumber \\
&=& \epsilon_{11} \sigma(e^- p \to N_1 j+X)/|(V_{lN})_{11}|^2.
\end{eqnarray}
Analogously the relations of Eqs.(\ref{sigma-ratio}) and
(\ref{sum-sigma}) are also satisfied by the production processes
$e^- p \to E_i^- j+X$ with the nearly degenerate heavy charged
lepton $E^-_i$.

From Fig.\ref{fig1} and Fig.\ref{fig2}, one can read off the cross
sections for $N$ and $E$ productions with appropriate constraints.
Eq.(\ref{ee0nu}) gives the strongest constraint on the couplings.
Using that constraint literately, it is not possible to have large
enough events produced. However, note that the couplings in
Eq.(\ref{ee0nu}) involve a summation of terms not necessarily
positively defined. If one can reconstruct the $N_i$ mass
eigenstates which may be able to do with sufficient mass splitting,
the production cross sections will be proportional to the square of
the absolute value of each term.  There are chances of cancelation
between terms\cite{He:2009lm} and leave individual terms sizeable.
Such cancelation occurrence has to be determined by experiments. We
will discuss this more later. Here we will assume that this is the
case, and then take the upper bounds $\epsilon_{11} = 3\times
10^{-3}$, $3.6\times 10^{-4}$ for the ST-I and ST-III models,
respectively, for discussions. From the figures we see that for $E_e
= 70$ GeV, the cross section for $N$ production in the ST-I model,
can be as high as a few fb with mass as large as 700 GeV. For $E_e =
140$ GeV, even with a mass scale  as large as a TeV the cross
section can be more than a $fb$. These cross sections are much
higher than that can be achieved for single $N$ production at the
LHC. With an integrated luminosity of $100~fb^{-1}$, there are
several hundred events can be produced and analyzed. For the ST-III
model, the event number is smaller. But can still have hundreds of
events for the mass scale to be as high as 800 GeV with $E_e = 140$
GeV. In this case there is also the possibility of studying $E$
production. The mass scale can be probed up 800 GeV too.

\subsection{Signals of $N$ and $E$}

The identification of the productions of $N$ and $E$ can be studied
by reconstructing decay products of $N$ and $E$. In order to have
large enough production cross sections, the parameter
$(V_{lN})_{ij}$ need to be close to their allowed bounds. With these
bounds, $N$ and $E$ will decay inside the detector. The final states
will involve a lepton and a $W$ or $Z$ from $N$ or $E$ decays, and
it would be most convenient to choose the charged lepton in the
final states to analyze. To keep the analysis as simple as possible,
one can reconstruct the final $W$ and $Z$ using their hadronic decay
modes. This procedure will reduce the available events for analysis
depending on the branching ratios of $N$ and $E$ decays. We now
provide some details.

In the ST-I model, the dominant decay modes for $N$ are
\begin{eqnarray}
&&\Gamma(N_i\to \ell^-W^+)= \Gamma(N_i\to \ell^+W^-)={g^2\over
64\pi}|(V_{lN})_{\ell i}|^2{m_{N_i}^3\over m_W^2}\left(1-{m_W^2\over
m_{N_i}^2}\right)\left(1+{m_W^2\over m_{N_i}^2}-2{m_W^4\over
m_{N_i}^4}\right)\label{N1},\nonumber\\
&&\sum^3_{m=1}\Gamma(N_i\to \nu_mZ^0)={g^2\over 64\pi
c_W^2}\sum_{\ell=1}^3|(V_{lN})_{\ell i}|^2{m_{N_i}^3\over
m_Z^2}\left(1-{m_Z^2\over m_{N_i}^2}\right)\left(1+{m_Z^2\over
m_{N_i}^2}-2{m_Z^4\over
m_{N_i}^4}\right)\label{N2},\nonumber\\
&&\sum^3_{m=1}\Gamma(N_i\to \nu_m h^0)={g^2\over
64\pi}\sum_{\ell=1}^3|(V_{lN})_{\ell i}|^2{m_{N_i}^3\over
m_W^2}\left(1-{m_h^2\over m_{N_i}^2}\right)^2\label{N3}.
\end{eqnarray}
In the above equations, we have assumed that $V_{PMNS}$ is unitary.
For the ST-III model, the neutral heavy neutrino $N$ has the same
form as that in the ST-I model given above, while for the heavy
charged lepton $E$ decay widths are given by
\begin{eqnarray}
&&\sum^3_{m=1}\Gamma(E_i^+\to \bar{\nu}_mW^+)={g^2\over
32\pi}\sum_{\ell=1}^3|(V_{lN})_{\ell i}|^2{m_{E_i}^3\over m_W^2}\left(1-{m_W^2\over
m_{E_i}^2}\right)\left(1+{m_W^2\over m_{E_i}^2}-2{m_W^4\over
m_{E_i}^4}\right)\label{E1},\nonumber\\
&&\Gamma(E_i^+\to \ell^+Z^0)={g^2\over 64\pi c_W^2}|(V_{lN})_{\ell
i}|^2{m_{E_i}^3\over m_Z^2}\left(1-{m_Z^2\over
m_{E_i}^2}\right)\left(1+{m_Z^2\over m_{E_i}^2}-2{m_Z^4\over
m_{E_i}^4}\right)\label{E2},\nonumber\\
&&\Gamma(E_i^+\to \ell^+h^0)= {g^2\over 64\pi}|(V_{lN})_{\ell
i}|^2{m_{E_i}^3\over m_W^2}\left(1-{m_h^2\over
m_{E_i}^2}\right)^2.\label{E3}
\end{eqnarray}

\begin{figure}[htbp]
\vspace{5mm}
 \centering
 \includegraphics[width=10cm]{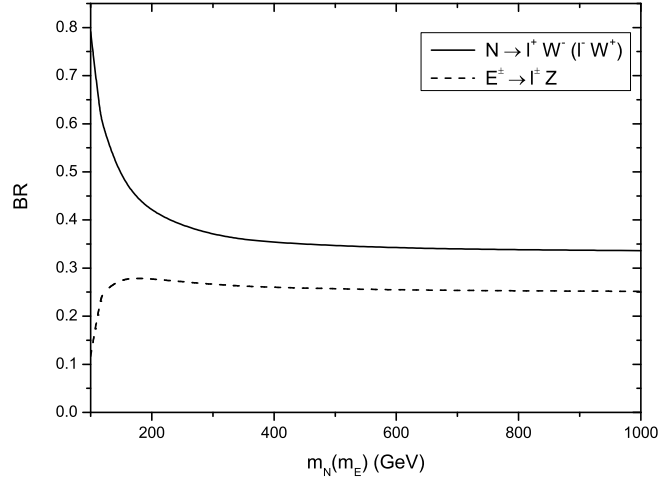}
\caption{\label{fig3} The branching ratios for $N \to l^\pm W^\mp$
and $E^\pm \to l^\pm Z$ as functions of $m_{N}$ and $m_E$. }
\end{figure}
In Fig.\ref{fig3}, we show the branching ratio for $N \to l^\pm
W^\mp$ and $E^\pm \to l^\pm Z$ as functions of heavy seesaw mass. We
see that the branching ratios are of order $33\%$ and $25\%$ when
the heavy neutrino and heavy charged lepton masses are large enough,
for example above 200 GeV.

Combining the hadronic decay branching ratios for $W^\mp$ and $Z$
are $67.6\%$ and $69.91\%$, we find that although there are
reduction factors, but there are still 22\% and 17\% of the produced
$N$ and $E$ can be analyzed by $N \to l^\pm + W^\mp$ and $E^{\pm}\to
l^{\pm} +Z$ followed by $W$ and $Z$ decay into hadronic final
states, respectively.

\subsection{The process $e^- p \to l^{\pm} W^{\mp}j+X$}

There are two types of processes, $e^- p \to l^+ W^-j+X$ and $e^- p
\to l^- W^+j+X$, which can be used to study the heavy neutrinos
$N_i~(i=1,2,3)$. The former type also contains information about the
Majorana nature of $N_i$.

After a straightforward calculation, we find that, for degenerate heavy neutrinos, the cross sections
for the processes $e^- p \to l^+ W^-j+X$ with $l =e,\;\mu,\;\tau$
are proportional to $|\kappa_{11}|^2$, $|\kappa_{12}|^2$ and
$|\kappa_{13}|^2$, respectively, where the $\kappa$ matrix is
defined as
\begin{eqnarray}
\kappa = V_{lN}^{*} V_{lN}^{\dag}\;.
\end{eqnarray}
Analogously, the cross sections for the processes $e^- p \to l^-
W^+j+X$ ($l =e,\;\mu,\;\tau$) are proportional to
$|\epsilon_{11}|^2$, $|\epsilon_{12}|^2$ and $|\epsilon_{13}|^2$
with the definition of
\begin{eqnarray}
\epsilon = V_{lN} V_{lN}^{\dag}\;.
\end{eqnarray}

Being a lepton number conserving process, $e^- p \to e^- W^+j+X$ has
large SM background, therefore, may be difficult to analyze. The
process $e^- p \to e^+ W^-j+X$ which violates lepton number by two
units may be more interesting to study. However, if the heavy
neutrinos are indeed nearly degenerate and one cannot separately
identify whether $e^+$ is from which $N_i$, the constraint from the
neutrinoless double-beta decay (Eq.(\ref{ee0nu})) implies that
\begin{eqnarray}
|\kappa_{11}| \,\,\le\,\,
5\times10^{-5}\times\frac{m_N}{1000\rm\,GeV}\;,
\end{eqnarray}
then the cross section for the process $e^- p \to e^+ W^-j+X$ is too
small to be measured. In this case the cross section for $e^- p \to e^+ W^-j+X$ is too small to be measured.

If the heavy neutrinos have sufficiently large mass splitting, one
can reconstruct the individual $N_i$ state by the invariant mass of
$e^+ W^-$, then one can have $e^- p \to e^+ W^-j+X$ with three
different invariant mass of $e^+ W$. The coherent cancelation in the
degenerate case would not happen with each of the cross section
proportional to the product of $|(V_{lN})_{1i}|^2$ and the $N_i$
branching ratio into $e^+$.  To this end we note that
$|(V_{lN})_{1i}|^2$ much larger than $|\kappa_{11}|$ is possible by
noticing the following: In the degenerate heavy neutrino case, the
constraint in Eq.(\ref{ee0nu}) is proportional to $\kappa_{11}$. But
for non-degenerate case, using $V_{lN} \hat{M}_N V^T_{lN}\approx 0$
the condition for large $V_{lN}$ solution, one can have solutions
such that the combination of $V_{lN}$ and $m_{N_i}$ in
Eq.(\ref{ee0nu}) is automatically equal to zero if $(m_{N_1}^2
-m_{N_2}^2)m_{N_3}/ (m_{N_1}^2 -m_{N_3}^2)m_{N_2} = -
(V_{lN})^2_{13}/(V_{lN})^2_{12}$. The mass splitting and the
couplings are therefore not separately constrained leaving the
possibility of large mass splitting and couplings. In this case, one
can then combine constraints $\epsilon_{11} < 3\times 10^{-3}$ and
$\epsilon_{11} < 3.6 \times 10^{-4}$, for ST-I nd ST-III,
respectively, to estimate the production cross section.

Another new class of lepton number violating processes which are
most promising to study at the LHeC are $e^- p \to l^{\pm}
W^{\mp}j+X$ ($l =\mu,\;\tau$) channels with $W$ decaying into hadron
jets $j_W$. All these processes violate generation lepton number.
For $l^+$ the lepton number is violated by two units. The detection
of this class of processes therefore can also reveal the Majorana
nature of the heavy leptons. Since the SM backgrounds for these
processes are small, they provide the opportunity to verify the
existence of heavy neutrinos $N_{1,2,3}$ cleanly without the
possible stringent constraint from neutrinoless double-beta decay
even the heavy neutrinos are degenerate.

Considering the fact that the upper bounds of $|\epsilon_{12}|$ and
$|\epsilon_{13}|$ are relatively large (see Eq.(\ref{eij}) for the
ST-I model and Eq.(\ref{fcnc'}) for the ST-III model),
the processes
\begin{eqnarray}
e^- p \to N_{1,2,3} + j + X \to l^- W^+ + j + X \to l^- j_W + j +
X\;,~~~(l =\mu,\;\tau)
\end{eqnarray}
at the LHeC may be practical.

The processes $e^- p \to N_{1,2,3} + j + X \to l^+ W^- + j + X \to
l^+ j_W + j + X\;,~~~(l =\mu,\;\tau)$ will be proportional to
$|\kappa_{12}|$ and $|\kappa_{13}|$. Unlike $\kappa_{11}$, there are
no cancelations in $\kappa_{12}$ and $\kappa_{13}$, in general.
Therefore, they can have the same order of magnitude as that for
$\epsilon_{12}$ and $\epsilon_{13}$ leading to similar cross
sections as that of $e^- p \to N_{1,2,3} + j + X \to l^- W^+ + j + X
\to l^- j_W + j + X$. In the following we will concentrate on $l^-$
cases.

If we assume that the heavy neutrinos of three generations are
quasi-degenerated and have the same mass ($m_N$) and decay width, we
get the relation of $\sigma(e^- p \to \mu^- W^+j+X): \sigma(e^-p \to
\tau^- W^+j+X) = |\epsilon_{12}|^2 : |\epsilon_{13}|^2$. In our
numerical calculations, we take the upper bounds $|\epsilon_{12}|=
1\times10^{-4}$, $|\epsilon_{13}| = 0.01$ for the ST-I model and
$|\epsilon_{12}| = 1.7\times10^{-7}$, $|\epsilon_{13}| = 4.2\times
10^{-4}$ for the ST-III model, respectively.

In Table \ref{tab2}, We list some of the numerical results of the
cross sections for the process $e^- p \to \tau^- W^+j+X$ and total
decay widths of heavy neutrinos with different values of $m_N$ in
both the ST-I and ST-III models. The total decay widths of heavy
neutrinos could be obtained from their dominant decay modes:
$N_{1,2,3} \to l^+ W^-$, $l^- W^+$, $\nu Z$, $\bar{\nu} Z$, $\nu h$,
$\bar{\nu} h$. We can see from the table that the cross section for
the $e^- p \to \tau^- W^+ j+X$ process can reach few hundreds of
$fb$ when the heavy neutrino mass is several hundred GeV. Since
$|\epsilon_{12}|^2 / |\epsilon_{13}|^2 \sim 10^{-4}-10^{-7}$, and
$\sigma(e^-p \to \tau^- W^+j+X) <2.02 \times 10^{3}~fb$ (see Table
\ref{tab2}), we obtain that the production cross sections for $e^- p
\to \mu^- W^+j+X$ process in the ST-I and ST-III models are
$10^{-4}$ and $10^{-7}$ smaller than for the $\tau^- W^+j$
production process respectively, and
\begin{eqnarray}
\sigma(e^- p \to \mu^- W^+j+X) = \sigma(e^- p \to \tau^- W^+j+X)
\times \frac{|\epsilon_{12}|^2}{|\epsilon_{13}|^2} <2.02 \times
10^{-1} fb \;.
\end{eqnarray}
Therefore, we can conclude that $e^- p \to \tau^- W^+j+X$ is an
ideal signal process for search the heavy neutrinos, but the $e^- p
\to \mu^- W^+j+X$ process is not, due to its small cross section.


\begin{table}[tp]
\begin{center}
\scalebox{0.85}{
\begin{tabular}{|c|c|c|c|c|c|c|} \hline
   &\multicolumn{3}{c|}{ST-I model} &\multicolumn{3}{c|}{ST-III model} \\ \cline{2-7}
 $m_{N}$&$\Gamma(N)$ &\multicolumn{2}{c|}{$\sigma(e^{-}p \to \tau^- W^+j+X)~(fb)$ } & $\Gamma(N)$&
     \multicolumn{2}{c|}{$\sigma(e^{-}p \to \tau^- W^+j+X) ~(fb)$ } \\ \cline{3-4} \cline{6-7}
  $({\rm GeV})$&$({\rm GeV})$ & $E_e=70{\rm GeV}$ & $E_e=140{\rm GeV}$
&$({\rm GeV})$& $E_e=70{\rm GeV}$ & $E_e=140{\rm GeV}$
\\ \hline
 100 & $1.017\times 10^{-3}$ & 1413 & 2017 &$1.132\times 10^{-4}$ & 22.40 & 31.98  \\
 200 & 0.05089 & 372.7 & 580.5 & 0.005663 & 5.901 & 9.199  \\
 400 & 0.5081 & 96.30  & 196.7 & 0.05655 & 1.530 & 3.114 \\
 600 & 1.773 & 25.08 & 80.29 & 0.1974 & 0.3952 & 1.2742 \\
 800 & 4.251 & 4.744 & 31.24 & 0.4731 & 0.07312 & 0.4928 \\
 1000& 8.345 & 0.5709 & 10.67 & 0.9287 & 0.00689 & 0.1645 \\
 \hline
\end{tabular}}
\end{center}
\begin{center}
\begin{minipage}{14cm}
\caption{\label{tab2} The numerical results of the cross sections
for the process $e^- p \to \tau^-W^+j+X$ and the total decay widths
of the heavy neutrinos by taking $|\epsilon_{13}| = 0.01$ for the
ST-I model and $|\epsilon_{13}| = 4.2\times 10^{-4}$ for the ST-III
model, respectively. $\Gamma(N)$ and $\sigma$ represent the decay
width of $N$ and the cross section for the process $e^- p \to
\tau^-W^+j+X$, respectively. There we take $E_p=7~{\rm TeV}$. }
\end{minipage}
\end{center}
\end{table}

\par
\begin{figure}[htbp]
\vspace{5mm}
 \centering
 \includegraphics[width=10cm]{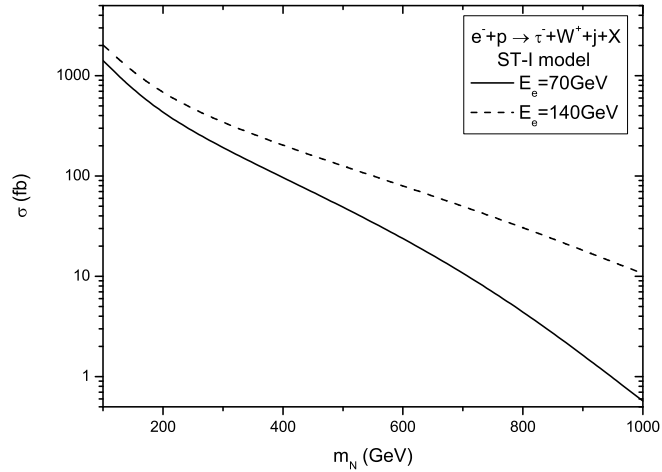}
\caption{\label{fig4} The cross sections for the $\tau^- W^+ j$
production in the ST-I model at the LHeC as functions of $m_{N}$
with $E_p=7~{\rm TeV}$ and $|\epsilon_{13}| = 0.01$, where the
full-line is for $E_e=70~{\rm GeV}$ and the dashed-line for
$E_e=140~{\rm GeV}$. }
\end{figure}
\begin{figure}[htbp]
\vspace{5mm}
 \centering
 \includegraphics[width=10cm]{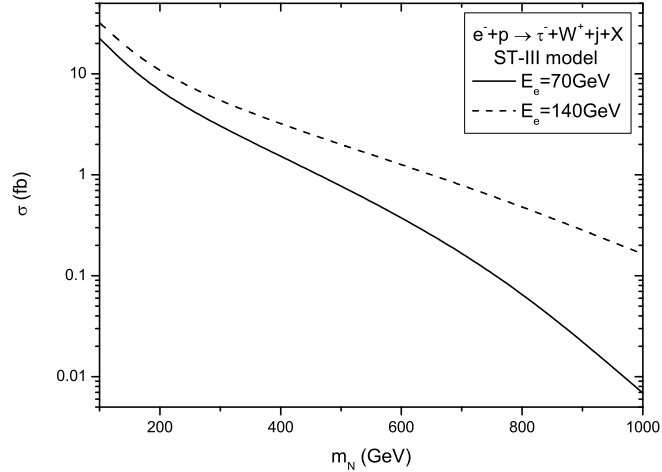}
\caption{\label{fig5} The cross sections for the $\tau^- W^+j$
production in the ST-III model at the LHeC as functions of $m_{N}$
with $E_p=7~{\rm TeV}$ and $|\epsilon_{13}| = 4.2\times 10^{-4}$,
where the full-line is for $E_e=70~{\rm GeV}$ and the dashed-line
for $E_e=140~{\rm GeV}$.  }
\end{figure}

In Fig.\ref{fig4} and Fig.\ref{fig5}, we depict the cross sections
for the process $e^-p \to \tau^- W^+j+X$ at the LHeC as functions of
the mass of $N$ in the ST-I and ST-III models, respectively. For
$E_e = 140$ GeV in the ST-I model, with $100~fb^{-1}$ integrated
luminosity, there are over one thousand events for the lepton number
violating process study for a seesaw mass scale of 1 TeV. Lowering
$E_e$ to 70 GeV, the reach of the seesaw mass scale is also lowered,
but can still have more than a thousand events to be studied for
seesaw mass scalae as high as 700 GeV. For the ST-III model, more
than a hundred events will be accumulated when $m_N < 700$ and 400
GeV for the case of $E_e = 140$ and 70 GeV, respectively.

\subsection{The process $e^- p \to l^{-} Z j+X$}

Analogous to the $N$ production, $e^- p \to E_{1,2,3}^- + j+X \to
l^-Z+j + X \to l^- j_{Z} + j + X$ with $l =\mu,\;\tau$ are lepton
number violating processes, therefore, might be signal processes for
the search of the heavy charged leptons $E_i$ ($i =1,\;2,\;3$). But
the $e^- p \to E_{1,2,3}^- j+X \to e^-Z+j + X \to e^- j_{Z} + j + X$
channel can not be an ideal signal process for testing the type-III
seasaw model because of the large SM background.

\begin{figure}[htbp]
\vspace{5mm}
 \centering
 \includegraphics[width=10cm]{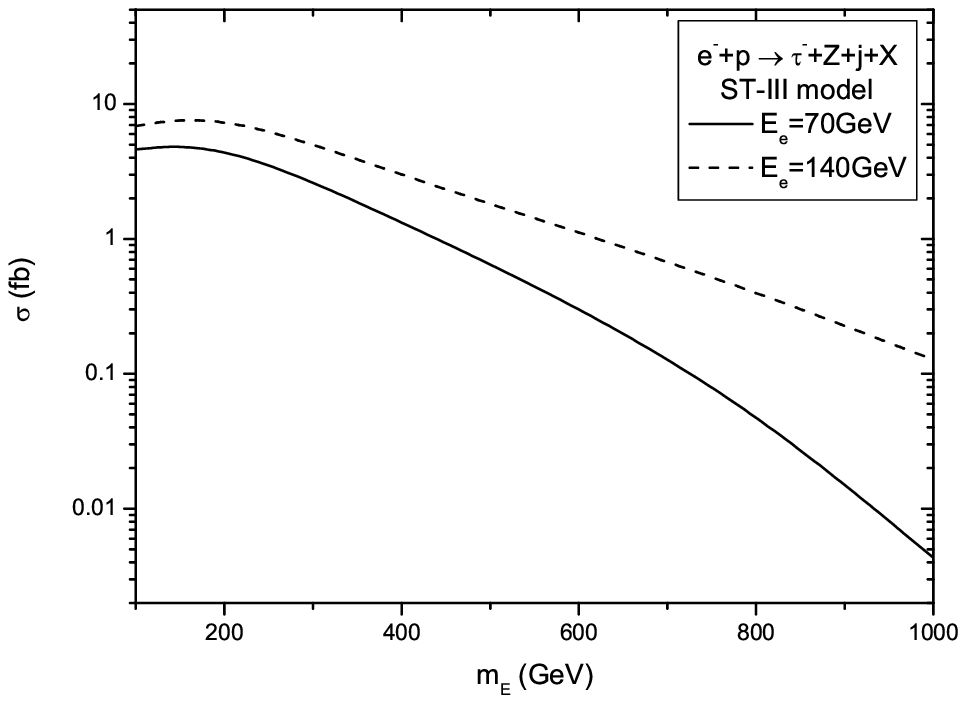}
\caption{\label{fig6} The cross sections for the $\tau^- Zj$
production in the ST-III model at the LHeC as functions of $m_{E}$
with $E_p=7~{\rm TeV}$ and $|\epsilon_{13}| = 4.2\times 10^{-4}$,
where the full-line is for $E_e=70~{\rm GeV}$ and the dashed-line
for $E_e=140~{\rm GeV}$.  }
\end{figure}

Since the experimental upper bound for $|\epsilon_{12}|$ is $1.7
\times 10^{-7}$ which is much smaller than that for
$|\epsilon_{13}|$ in the ST-III model, only the $\tau^- Zj$
production rate with interchanging heavy charged leptons $e^- p \to
E_{1,2,3}^- j+X \to \tau^-Zj + X $ is significant. We assume all the
heavy charged leptons of three generations are quasi-degenerated and
have the same mass ($m_E$) and decay width, and take
$|\epsilon_{13}| = 4.2\times 10^{-4}$ in the ST-III model. In
Fig.\ref{fig6} we present the $m_E$ dependence of the cross section
for the $\tau^- Zj$ production at the LHeC, where the full-line is
for $E_e=70~{\rm GeV}$ and the dashed-line for $E_e=140~{\rm GeV}$.
Some of the numerical results in Fig.\ref{fig6} are listed in Table
\ref{tab3}. We can see from these results that the cross section for
the $e^- p \to E^-_{1,2,3}j \to \tau^- Z j+X$ process can reach a
few $fb$ when the heavy charged lepton mass $m_E <600$ and 400 GeV
for $E_e=140$ and 70 GeV, respectively.
\begin{table}[tp]
\begin{center}
\scalebox{1}{
\begin{tabular}{|c|c|c|c|} \hline
 $m_{E}$&$\Gamma(E)$ &\multicolumn{2}{c|}{$\sigma(e^{-}p \to \tau^- Zj+X)~(fb)$ }  \\ \cline{3-4}
  (GeV)&$({\rm GeV})$ & $E_e=70{\rm GeV}$ & $E_e=140{\rm GeV}$   \\ \hline
 100 & $1.014\times 10^{-4}$ & 4.622 & 6.877  \\
 200 & 0.003993 & 5.297 & 8.731   \\
 400 & 0.03822 & 1.337  & 2.944   \\
 600 & 0.1324 & 0.3187 & 1.126    \\
 800 & 0.3165 & 0.05343 & 0.4077   \\
 1000& 0.6205 & 0.00435 & 0.1270  \\
\hline
\end{tabular}}
\end{center}
\begin{center}
\begin{minipage}{14cm}
\caption{\label{tab3} The numerical results of the cross sections
for the process $e^- p \to \tau^-Zj+X$ and the total decay widths of
the heavy charged leptons in the ST-III model. $\Gamma(E)$ and
$\sigma$ represent the decay width of $E$ and the cross section for
the process $e^- p \to \tau^-Zj+X$, respectively. There we take
$E_p=7~{\rm TeV}$ and $|\epsilon_{13}| = 4.2\times 10^{-4}$. }
\end{minipage}
\end{center}
\end{table}

\section{Conclusions}

We have studied the potential of testing the seesaw type I and III
models at the LHeC. The e-p collision mode provides an excellent
place to study lepton number violating processes $e^- p\to N_{i} j
+X \to e^+ W^-j+ X$ and $e^- p\to N_{i} j +X \to \tau^-W^+j+ X$ with
$W$ into hadron jets. Here $N_{1,2,3}$ are heavy Majorana neutrinos
and $j$ is a hard hadron jet. Although the process $e^- p\to N_{i} j
+X \to e^+ W^-j+ X$ is stringently constrained from neutrinoless
double-beta decay, there are solutions where this constraint can be
satisfied with sizeable production cross section. For the process
$e^- p\to N_{i} j +X \to \tau^{\pm}W^{\mp}j+ X$, the neutrinoless
double-beta decay constraint does not apply. With $E_e = 140$ GeV
and $E_p = 7$ TeV, we find that the cross section for the heavy
neutrino $N$ can be higher than what can be achieved by the p-p
collision mode of LHC with same related heavy neutrino couplings.
For $E_e = 140$ GeV, with $100~fb^{-1}$ integrated luminosity, there
are over one thousand events for the lepton number violating process
study for a seesaw mass scale of 1 TeV. Lowering $E_e$ to 70 GeV,
the reach of the seesaw mass scale is also lowered, but can still
have more than a thousand events to be studied for seesaw mass
scalae as high as 700 GeV. For the ST-III model, more than a hundred
events will be accumulated when $m_N < 700$ and $m_N < 400$ GeV for
the case of $E_e = 140$ and $E_e = 70$ GeV, respectively. With e-p
collision mode we can use the process $e^- p \to E_{1,2,3}^- j+X \to
\tau^-Zj + X $ with $Z$ into hadron jets to study the heavy charged
leptons in the ST-III model, whose cross section can reach a few
$fb$ when the heavy charged lepton mass $m_E < 600$ and 400 GeV for
$E_e=140$ and 70 GeV, respectively.

\acknowledgments This work was supported in part by the National
Natural Science Foundation of China(No.\-10875112, No.10675110), and
the Specialized Research Fund for the Doctoral Program of Higher
Education(No.20093402110030), and in part by NSC and NCTS.

\end{document}